\documentclass[%
 reprint,
 amsmath,amssymb,
 aps,
prb,
]{revtex4-1}

\usepackage{graphicx}      % Include figure files
\usepackage{bm}             	% bold math
\usepackage{times}			% Times font
\usepackage{tabularx}
\usepackage{color}
\newcommand{\IEF}{Institut d'Electronique Fondamentale, CNRS, Univ. Paris-Sud, Universit\'e Paris-Saclay, 91405 Orsay, France}
\newcommand{\IMEC}{Imec, Kapeldreef 75, B-3001 Leuven, Belgium}
\newcommand{\KUL}{KU Leuven, Departement Electrotechniek (ESAT),  Kasteelpark Arenberg 10, B-3001 Leuven, Belgium}

\begin{document}
\title{All electrical propagating spin wave spectroscopy with broadband wavevector capability}

\author{F. Ciubotaru} \email{Florin.Ciubotaru@imec.be} \affiliation{\IMEC}\affiliation{\KUL}
\author{T. Devolder} \affiliation{\IEF}
\author{M. Manfrini} \affiliation{\IMEC}
\author{C. Adelmann} \affiliation{\IMEC}
\author{I. Radu} \affiliation{\IMEC}
%\author{A. Thean} \affiliation{\IMEC}

\date{\today}                                           
%%%%%%%%%%%%%%%%%%%%%%%%%%%%%%%%%%%%%%%%
%
%       Abstract
%
%%%%%%%%%%%%%%%%%%%%%%%%%%%%%%%%%%%%%%%%
\begin{abstract}
We develop an all electrical experiment to perform the broadband phase-resolved spectroscopy of propagating spin waves in micrometer sized thin magnetic stripes. The magnetostatic surface spin waves are excited and detected by scaled down to 125 nm wide inductive antennas, which award ultra broadband wavevector capability. The wavevector selection can be done by applying an excitation frequency above the ferromagnetic resonance. Wavevector demultiplexing is done at the spin wave detector thanks to the rotation of the spin wave phase upon propagation. A simple model accounts for the main features of the apparatus transfer functions. Our approach opens an avenue for the all electrical study of wavevector-dependent spin wave properties including dispersion spectra or non-reciprocal propagation. % and wavevector dependent propagation loss. 
\end{abstract}

\maketitle

%%%%%%%%%%%%%%%%%%%%%%%%%%%%%%%%%%%%%%%%
%
%                Paper
%
%%%%%%%%%%%%%%%%%%%%%%%%%%%%%%%%%%%%%%%%
%\section{Introduction}
Spin wave based computing \cite{chumak_magnon_2015}-- a paradigm-shifting technology that uses the interference of spin waves-- offers potential for significant power and area reduction per computing throughput with respect to complementary metal-oxide-semiconductor (CMOS) transistor technology. Efficient solutions for spin wave routing\cite{vogt_spin_2012, vogt_realization_2014}, spin wave emission \cite{demidov_direct_2010}, amplification \cite{chumak_magnon_2015}, and spin wave combination \cite{rousseau_realization_2015, klingler_design_2014} have been developed. However, these solutions often rely on  materials that are difficult to integrate \cite{cherepanov_saga_1993} into a CMOS environment. Moreover,  they are often demonstrated only for long wavelength $(\geq 1~\mu \textrm{m})$ spin waves, for which the low group velocity limits the speed of computation and communication. Efficient methods to generate and detect spin waves with short wavelengths are still lacking. Inductive methods have commonly been employed for long wavelength spin waves as has Brillouin light scattering spectroscopy\cite{demokritov_brillouin_2001}, which is however diffraction limited and requires complex procedure to retrieve phase information \cite{fohr_phase_2009}. Alternative spin wave generation and detection methods based on magneto-elastic coupling in surface-acoustic wave devices \cite{cherepov_electric-field-induced_2014} are still under development  and raise questions regarding their high frequency capability \cite{jegou_development_2014}. Long-to-short wavelength conversion can be done in magnonic crystals \cite{yu_omnidirectional_2013} for a geometrically limited discrete set of wavevectors at the expense of high conversion loss. A better conversion efficiency can be obtained by periodically folded coplanar antennas \cite{vlaminck_current-induced_2008, vlaminck_spin-wave_2010} but this at the expense of any flexibility in the generated wavevevector. Overall, none of the above methods has so far  demonstrated the combination of phase resolution, broad frequency coverage, high sensitivity, and large wavevector (short wavelength) capability.

In this work, we demonstrate that the use of deep sub-micron inductive antennas can circumvent these limitations and allow for the generation and detection of spin waves with ultra-wide frequency band capability and broad wavevector capability up to $15-20~\textrm{rad}/\mu \textrm{m}$. We illustrate our method on micrometer-sized Permalloy stripes and describe its behavior within an analytic framework. Our method complements the advantages of Brillouin Light scattering in a compact all electrical device in which the phase resolution is inherent and the operation is convenient as it benefits from the speed -- and sensitivity -- of electronic measurement systems. Our geometry paves the way for all electrical characterization of wavevector-dependent properties of spin waves like chiral effects \cite{belmeguenai_interfacial_2015}, magnon-magnon interactions and spin current to spin wave interactions \cite{vlaminck_current-induced_2008} that need to be understood to assess the technological potential of spin wave based computing.

%%%%%%%%%%%%%%%%%%%%%%%%%%%%%%%%%%%%%%%%
%\section{Samples and set-up}
Our devices are based on sputter-deposited $\textrm{Ta~3~nm} / \textrm{Ni}_{80}\textrm{Fe}_{20}$ (Permalloy), $t=17~\textrm{nm}/ \textrm{Ta~3~nm} $ films grown on top of a 300 nm layer of $\textrm{Si}\textrm{O}_{2}$ on Si substrate.  Their hysteresis loops indicated a very soft and quasi-isotropic behavior, consistent with a polycrystalline and not a textured structure. A saturation magnetization of 771 kA/m and a damping of 0.008 was extracted from ferromagnetic resonance (FMR) experiments. The Permalloy layer was patterned into stripe-shaped spin wave conduits with widths in the range of $w \in [0.25,~5~\mu\textrm{m}]$ using  electron beam lithography and standard ion-beam patterning. The conduits were then covered by 100~nm of hydrogen silsesquioxane (HSQ) for electric isolation.  

\begin{figure}
\includegraphics[width=8.5cm]{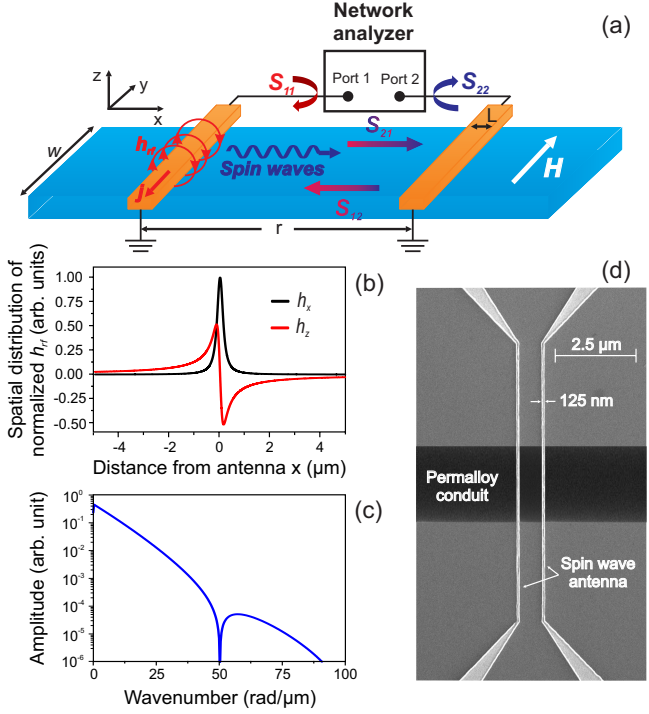}
\caption{(Color online) Experimental configuration: (a) Schematic of the studied devices consisting of inductive antennas (orange) and a magnetic stripe conduit (blue) with propagating spin waves (wavy arrows).(b) Spatial distribution of the $h_x$ and $h_z$ components of the magnetic fields induced by the microwave current flowing through a 125 nm wide antenna placed at 100 nm above the permalloy stripe. (c) Wavenumber distribution of the $h_x$ field in (b). (d) Scanning electron micrograph of a 2.5 $\mu$m wide Permalloy conduit with two 125 nm antennas. }
\label{FIG_CONFIG}
\end{figure}

Spin wave excitation and detection was performed via two identical Ti/Au antennas with widths in the range of $L \in [0.125,~1 ~\mu \textrm m]$ and  center-to-center distances ranging from 1.2 to 10 $\mu$m, denoted below as $r$. Each antenna is connected in series between a ground pad and another coplanar contact pad (Fig.~\ref{FIG_CONFIG}).  The antennas are RF powered and thus generate an oscillating RF field $h_x$ located essentially under the antennas and a weaker and anti-symmetric RF field $h_z$ located at the edges of the antennas [see example in Fig.~\ref{FIG_CONFIG}(b)]. In this geometry, the RF fields will thus excite spin waves with wavevectors purely along the $x$ direction. %Note that the spin waves with wavevectors perpendicular to the film plane  - the so called perpendicular standing spin wave modes (PSSW) - have frequencies above our region of investigation, therefore, we shall not consider them.

Static longitudinal fields saturating the magnetization along the stripe length lead to signals related to spin waves that are excited by  $h_z$  at the antenna edges. However the associated signals  (not shown) were one to two orders of magnitude smaller than when a transverse magnetic field was applied in the configuration depicted in Fig.~\ref{FIG_CONFIG}(a). In the remainder of the paper, we focus on the case of a static field $H_y$ saturating the stripe magnetization in the transverse direction (Damon-Eshbach configuration), and thus study the properties of magnetostatic surface spin waves (MSSW) that are excited in this geometry.
%%
%	Figure
%%
%

The devices were characterized by measuring the magnetic field dependence of their complex scattering matrix $S_{i,j}$ with $i, j  \in \{1, 2\}$ under 10 mW of excitation power. An on-chip full two-port calibration routine with a load-match-reflect standard impedance calibration kit was performed to correct for imperfections of the network analyzer, the cable assembly, and the RF probes. The pads connecting the device were short enough (300 $\mu$m) that no deembedding was required. The phase error associated with this approximation is negligible compared to the phase accumulation due to propagation of spin waves through our devices, as shown below. \\
The signal transmission through the device comprises two superimposed signals: (i) a direct antenna-to-antenna parasitic coupling of typically -30 to -40 dB and (ii) smaller superimposed spin wave related signals. We construct an approximation of the scattering matrix of the parasitic transmission by calculating the average of the scattering matrix over all applied magnetic fields. We then isolate the spin-wave related signals by defining the complex admittance normalized to $1 / 50~\Omega$ as:

\begin{equation}
\label{admittance}
\tilde Y(H_y) = {\frac{1}{2}} \Big[ \tilde S (H_y) - \frac{1}{2H_\textrm{max}}\int_{-H_\textrm{max}}^{H_\textrm{max}} \tilde S(H_y)\, \mathrm{d}H_y \Big]
\end{equation}

The typical order of magnitude of $|| \tilde Y ||$ is $10^{-5}$ for $r=1.1~\mu\textrm m$, $L=125~\textrm{nm}$ and $w=2.5~\mu\textrm m$, and seems to depend on the spin wave conduit width.

\begin{figure*}
\includegraphics[width=17cm]{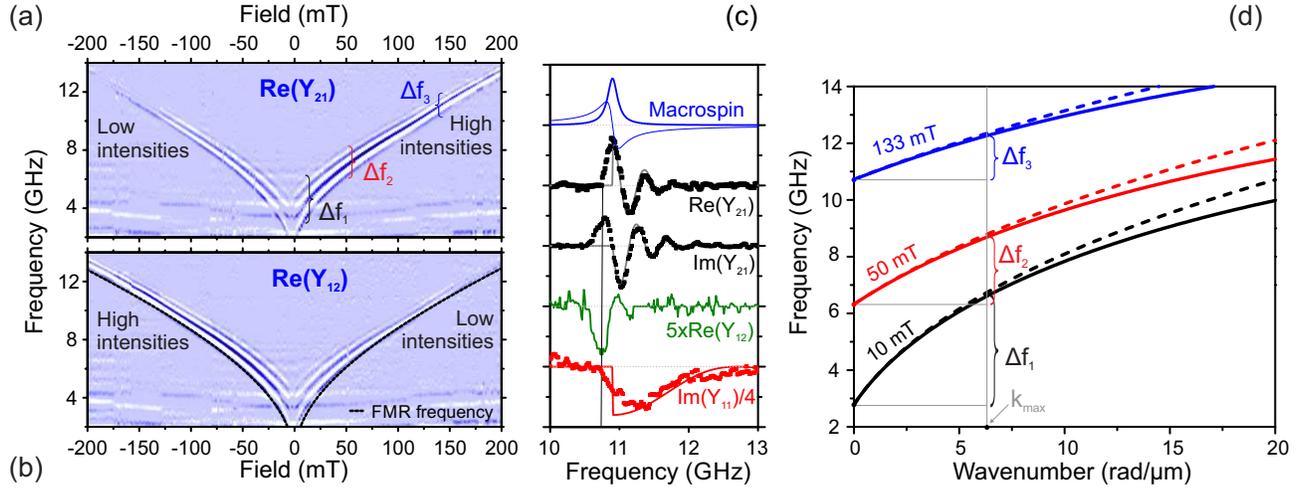}
\caption{(Color online) Field dependence of the real parts of the forward transmission (a) and backward transmission (b) signals (Eq.~\ref{admittance}) for a Permalloy stripe with $w=5~\mu\textrm m$, a propagation distance of $r=4~\mu\textrm m$, and antenna widths of $L=1~\mu\textrm m$. Light blue color corresponds to a zero spin-wave transmission, while the dark blue and the white stands for propagating spin waves. The dashed black curve in (b) is the analytical field-dependency of the ferromagnetic resonance frequency. (c) Comparison of the forward transmission (black symbols), backward transmission (green symbols) and forward reflection (red symbols) for a field of 140 mT along the stripe width. The latter have been rescaled for readability. The lines superimposed on the experimental response curve are the responses modeled according to Eq.~\ref{S21} with $\beta=0.15$. The top curves are the theoretical macrospin response using material parameters determined by FMR. (d) Spin wave dispersion relations calculated analytically for the structure in (a) with (dashed curves) and without (continuous lines) taking into account the exchange interactions for three values of the applied magnetic field as indicated. $k_{max} = 6.28~\textrm rad/\mu\textrm m$ is the maximum spin-wave wavevector that can be excited by a $1~\mu\textrm m$ wide antenna.}
\label{NonReciprocityFigure}
\end{figure*}

%%%%%%%%%%%%%%%%%%%%%%%%%%%%%%%%%%%%%%%%
%\section{Results}
%\subsection{Main experimental features}
%\textbf{Message 1} (Raw description of the results, explananation of the overall shape: threshold).
The basic behavior of the devices is illustrated in Figs.~\ref{NonReciprocityFigure}(a-c), which display the transmission and reflection properties for a stripe width of $w=5~\mu\textrm m$, a propagation distance of $r=4~\mu\textrm m$, and antenna widths of $L=1~\mu\textrm m$. The field-independent cross-talk is subtracted using Eq.~\ref{admittance}. For a given magnetic field, the spin-wave transmission signal exists only above a certain frequency threshold and stays detectable within a frequency transmission band of width $\Delta f$. In all cases, the onset of finite transmission comes together with a drop of the reflection signal [red line in Fig.~\ref{NonReciprocityFigure}(c)]. \\ % threshold :
The frequency threshold shifts to higher frequencies upon increasing the magnetic field strength. We have found that it matches quantitatively the FMR condition $\omega_\textrm{FMR} \approx \gamma_0 \sqrt{H_y (H_y + M_S) }$ [see Fig.~\ref{NonReciprocityFigure}(b)]. Here, $\gamma_0$ is the gyromagnetic ratio and the shape anisotropy of the stripe has been neglected. \\ % transmission band :
Conversely, the transmission band $[f_\textrm{FMR}, f_\textrm{FMR}+\Delta f]$ gets narrower as we increase the field strength. This behavior can be understood by looking at the expected dispersion relation of spin waves\cite{Kalinikos} for several magnetic fields [see e.g. Fig.~\ref{NonReciprocityFigure}(d)].  The narrowing of transmission band stems from the fact that the antenna can excite spin waves only up to a finite wavenumber $k_\textrm{max}$. The flattening of the dispersion relations at larger magnetic fields [Fig.~\ref{NonReciprocityFigure}(d)] then leads to a narrowing of the transmission band $\Delta f = f(k_\textrm{max})-f(k=0)$ when the magnetic field is increased. 

%\textcolor{red}{ what do you mean in the next sentence ? Additionally, a further narrowing of the transmission band $\Delta f$ observed in experiments}. We will see that it is related to a decrease of the excitation efficiency for spin waves with shorter wavelength. 
%To note that the antennas used in the experiments presented in Fig.~\ref{NonReciprocityFigure} have an width of $L=1~\mu\textrm m$, therefore, they can excite spin waves with wavenumbers up to $k_{max} = 6.28~\textrm rad/\mu\textrm m$. Nevertheless, spin waves with even a broader band of wavenumbers can be excited by using an antenna with a much smaller width. For example, a 125 nm wide antenna can potentially excite spin waves having wavenumbers up to 50 rad/$\mu\textrm m$ (Fig.~\ref{FIG_CONFIG}(c)). A description of the experimental results obtained for the later antenna size will follow up later in the paper.

%\textbf{Message 2} (propagation leads to dephasing = phase rotation of the spin wave).
The shapes of the reflection and transmission signals [Fig.~\ref{NonReciprocityFigure}(c)] differ substantially: the transmission signals are complex numbers with real and imaginary parts that oscillate and decay with frequency, while the reflection signals do not oscillate and show a slightly less pronounced decay with frequency. The oscillation of the transmission signal is thus a phase rotation inherent to the propagation of the spin waves. The decay of the envelope of the reflection and transmission signals stems from the wavevector-dependent excitation efficiency. The faster decay of the propagation signals may be indicative of additional losses upon spin wave propagation. \\
%\textbf{Message 3} (non reciprocity. Field polarity addressable.) 
A final striking point is the amplitude non reciprocity. For positive fields, the forward transmission is typically five times larger than the backward transmission, while the situation is reversed upon a change of the applied field direction (Fig.~\ref{NonReciprocityFigure}). This feature recalls the surface character of the MSSW of finite wavevector, and it will be discussed further below.

%%
%	Figure
%%
%

%%%%%%%%%%%%%%%%%%%%%%%%%%%%%%%%%%%%%%%%%%%%%%%%%%%%%%%%%%%%%%%%%%%%%%%%%%%%%%%%
%\subsection{Qualitative model describing the main features}
$ $ 

The main experimental features -- frequency threshold to allow transmission, phase rotation upon transmission, and non reciprocity -- can be accounted for by the following simple model.  
The pumping field has its components and its gradients along the stripe axis $x$ and stripe normal $z$ [Fig.~\ref{FIG_CONFIG}(a)].  The $z$ dependence is due to RF absorption by Eddy currents and excitation of the (lossy) magnetic precession \cite{bailleul_shielding_2013}, and it has its importance since MSSW waves tend to localize at either the top or the bottom surface of a ferromagnetic film \cite{damon_magnetostatic_1961, kostylev_non-reciprocity_2013}, depending whether ${k_x}$, \textbf{M} and $z$ form a direct or an indirect trihedron. %\textcolor{red}{It is correct}. 
The demagnetizing field within the stripe is quasi-uniform, except for a $y$ gradient of its $y$ component near the stripe edges. We have performed micromagnetic simulations\cite{OOMMF} and have found that a transverse bias field of 10 mT is enough to saturate the sample over $75\%$ of its width. % \textcolor{red}{that will be responsible for edge modes, to be considered later}.
All together, this provides in principle possibilities of exciting spin waves with propagating $k_x$ and $k_z$ components and with standing waves with $k_y$ near the edges.

Since spin waves with wavevectors $k_z$ perpendicular to the film plane have frequencies substantially above our region of investigation, we shall restrict our study to the case of $k_z=0$. The main transmission channel relies on spin waves extending over the whole stripe width, so only spin waves with wavevectors  purely along $x$ (Damon-Eshbach modes \cite{damon_magnetostatic_1961}) will be included in the model. For long wavelength (i.e. $k_x t << 1$) spin waves of frequency $\omega / (2\pi)$ the real part of the complex wavevector can be approximated by an exchange-free formulation:
%###################################
\begin{equation}
{k}_x = \frac{2}{t} \frac{\omega^2-\omega_\textrm{FMR}^2}{\gamma_0^2 M_S^2} \mathcal{H}(\omega-\omega_\textrm{FMR})
\label{DispersionRelation}
\end{equation}
%###################################
% I decided to drop the real parts
where $\mathcal{H}$ is the Heaviside distribution. From Fig.~\ref{NonReciprocityFigure}(d) one can observe that there is nearly no difference between the dispersion relations calculated with and without taking into account the exchange interaction in the studied wavevector range. \\

We approximate the Oersted field profile below the antenna by assuming that its $x$-component $h_\textrm{max}(z)$ is uniform in an interval $[-L_\textrm{eff}/{2}, L_\textrm{eff}/{2}]$  and vanishes everywhere else. $L_\textrm{eff} \approx {L+h}$ is the typical lateral extension of the RF field at the stripe altitude. The description is accurate only when $L \geq h$ [Fig.~\ref{FIG_CONFIG}(b)] but we shall see that the details beyond the lateral extension of the RF field are not critical. Equivalently, the RF field can be described in reciprocal space by:
\begin{equation}
h_x \propto \textrm{sinc} \Big[\frac{k_x L_\textrm{eff}}{2}\Big] 
\label{Sinc}
\end{equation}
% Thibaut's convention is that sinc(x) = (sin(x)) / x
This field profile changes sign at a wavevector of $2 \pi / L_\textrm{eff}$: our excitation efficiency vanishes at this wavevector and its multiples [Fig.~\ref{FIG_CONFIG}(c)]. 

We shall account for the overlap between the RF field profile and the exponential profile of the surface spinwaves by defining an ad-hoc $k_x$-dependent pumping (or pick-up) efficiency factor: 
\begin{equation}
\label{overlapfactor}
\chi = \frac{1}{t} \int_{-t/2}^{t/2} h_\textrm{max}(z) ~\textrm{exp}[- k_x z] \textrm{d}z
\end{equation}

If the center-to-center distance between the antennas is $r$, the spinwave undergoes a phase rotation of ${k}_x r$ along its propagation path, together with an exponential decay of rate $\beta {k}_x r$, where $\beta$ is a mode-dependent function of the Gilbert damping. This can be described by a propagation operator

\begin{equation}
\textrm{exp}[-i k_x r - \beta {k}_x r]
\label{Exp}
\end{equation}
Note that each maximum in the real part of the propagation operator corresponds to a total propagation over an integer number of wavelength. This will be a convenient way to directly measure $k_x$ from the oscillatory experimental signals.

Overall, the scattering admittance elements of our system at a given frequency are proportional to the product of the stripe width, the $k_x$ dependence of the excitation field (from Eq.~\ref{Sinc}) at the corresponding wavevector (from Eq.~\ref{DispersionRelation}), the detection sensitivity (also Eq.~\ref{Sinc}), the square of the pumping efficiency (Eq.~\ref{overlapfactor}) and the complex propagation operator (Eq.~\ref{Exp}).  
This yields:
\begin{equation}
Y_{i \neq j} \propto w  ~\chi^2 ~ \textrm{sinc}^2 \Big[\frac{k_x L_\textrm{eff}}{2}\Big] ~\textrm{exp}[-i k_x r - \beta {k}_x r]
\label{S21}
\end{equation}
Note that because of the surface nature of MSSWs, $\chi^2$ will vary faster than $\textrm{exp}[-2k_x t]$ and thus it will differ for forward $(S_{21})$ and backward $(S_{12})$ transmission coefficients. The difference increases with the wavevector, hence with frequency at a given applied magnetic field. This faster decay becomes clear in Fig.~\ref{NonReciprocityFigure}(c) when comparing forward (black) and backward (green) transmission signals. Note also that the transmission coefficients are complex numbers, with real and imaginary parts that rotate in quadrature thanks to the propagation operator. This is also in line with experiments. The reflection coefficients $S_{11}$ and $S_{22}$ can be obtained from Eq.~\ref{S21} by setting the propagation distance $r$ to zero. 

%%%%%%%%%%%%%%%%%%%%%%%%%%%%%%%%%%%%%%%
%\subsection{Comparison with experiments and discussion}

%what works
Equation~\ref{S21} is compared quantitatively to the experimental data in Fig.~\ref{admittance}(c). The main experimental features -- oscillations in quadrature of the two components of the transmission parameters and overall decay with the frequency -- are well reproduced with the propagation loss $\beta$ as single fitting parameter, which confirms that the essential physics is included in Eq.~\ref{S21}. 
%deficiencies

%%
%	Figure
%%
%
\begin{figure}
\includegraphics[width=8.3cm]{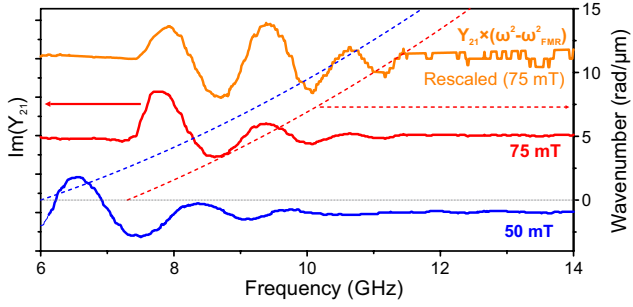}
\caption{(Color online) Examples of spin wave signals transmission (bold lines) and wavenumber (dashed lines, from Eq.~\ref{DispersionRelation}) for the narrowest antenna $L=125$ nm, the shortest propagation distance $r =1.125~\mu\textrm m$ and a stripe width of 5 $\mu$m. The top curve has been rescaled proportionally to the wavenumber to better evidence the signal oscillations at higher frequencies.} 
\label{NarrowAntenna}
\end{figure}

However, the onset of transmission near $\omega_\textrm{FMR}$ is much more gradual in the experiments than in the modeling. This shortfall of Eq.~\ref{S21} results from the assumption of a bijective relation between wavevector and frequency (Eq.~\ref{DispersionRelation}) while in reality any given mode has a finite susceptibility also below and above its eigenfrequency. For $k=0$ (FMR macrospin mode) the susceptibilities are symmetric and antisymmetric Lorentzian lines that show a finite spread around the mode center frequency (see Fig.~\ref{NonReciprocityFigure}(c) blue curves). The rigorous description of the onset of transmission at the threshold goes beyond the scope of this paper, but would require to include a convolution of the mode susceptibilities within the kernel of Eq.~\ref{S21}. One can notice that the experimental signals indeed resemble the convolution of the modeled response with a macrospin response [see Fig.~\ref{NonReciprocityFigure}(c)].

%wavevector capability
To be able to excite spin waves with an ultra-wide wavenumber band one must use antennas with smaller widths. For example, a 125 nm wide antenna can potentially excite spin waves with  wavenumbers up to 50 rad/$\mu\textrm m$ (Fig.~\ref{FIG_CONFIG}(c)). Figure~\ref{NarrowAntenna} demonstrates that spin waves can be efficiently generated and sensed inductively by a such a narrow antenna. For the shortest propagation distance of $r=1.125~\mu \textrm{m}$, few oscillations of the transmission signals were detected. The corresponding signals could be satisfactorily accounted for an effective $L_\textrm{eff}= 500$~nm (fits not shown) which corresponds to a first lobe of Eq.~\ref{Sinc} at $12~\textrm{rad}/\mu \textrm{m}$.  From the analysis of the dispersion relations calculated for different magnetic bias fields, we could estimate that the highest detected wavevector was $k_\textrm{h,30~\textrm{mT}} \approx  17~\textrm{rad}/\mu \textrm{m}$ (wavelength $\lambda \sim 370~\textrm{nm}$),  $k_\textrm{h,50~\textrm{mT}} \approx  15~\textrm{rad}/\mu \textrm{m}$ ($\lambda \sim 420~\textrm{nm}$) and  $k_\textrm{h,75~\textrm{mT}} \approx  12~\textrm{rad}/\mu \textrm{m}$ ($\lambda \sim 525~\textrm{nm}$), respectively. Our approach is very advantageous because it identifies the routes to access even higher  wavevectors. An improvement of the wavevector capability requires the minimization of the effective antenna size $L_\textrm{eff}$ by thinning the insulator that separates the antenna from the stripe, as well as a further shrinking of the antenna. Nevertheless, this will increase the parasitic cross-talk and thus a geometrical compromise will have to be found.

%For the shortest propagation distance of $r=1.125~\mu \textrm{m}$, 4 maxima (the FMR and 3 additional ones) of the transmission signals were detected. From the    such that the highest detected wavevector was $k_\textrm{high} \approx  17~\mu \textrm{m}^{-1}$. Our model is useful because it identifies the routes to access to further higher wavevectors. Improving the wavevector capability requires to minimize the effective antenna size $L_\textrm{eff}$ by thinning the insulator that separate the antenna from the stripe, and shrinking the antenna width further. 

%As this will increase parasitic cross-talk, a geometrical compromise will have to be found to avoid that operation at higher frequencies ends up being limited by the dynamic range of the network analyzer, as it is the case presently.

In summary, we have presented an inductive method to perform phase-resolved spectroscopy of propagating spin waves in thin Permalloy stripes. We demonstrate that very narrow antennas can be successfully used to generated and detect spin waves in a very wide range of wavevectors (up to $17~\textrm{rad}/\mu \textrm{m}$), which compares well with other techniques. Furthermore, we developed an analytic model to describe the main features of the spin-wave transmission functions as the dispersion relation or the non-reciprocity in propagation.

%The method was demonstrated in micrometer sized thin Permalloy stripes and was shown to operate till wavevectors of $17~\mu \textrm{m}^{-1}$, which compares well with other techniques. 

This work was supported by imec's Industrial Affiliation Program on Beyond CMOS devices and by the French National Research Agency (ANR) under contract No. ANR-11-BS10-0003 lead by Matthieu Bailleul. F.C. thanks J. Loo for e-beam lithography, Rudy Caluwaerts for SEM images and imec's clean room technical support.

%\bibliography{bib.bib}

\end{document}